# Steady state availability general equations of decision and sequential processes in Continuous Time Markov Chain models

E. M. Vasconcelos[1]

Continuous Time Markov Chain (CMTC) is widely used to describe and analyze systems in several knowledge areas. Steady state availability is one important analysis that can be made through Markov chain formalism that allows researchers generate equations for several purposes, such as channel capacity estimation in wireless networks as well as system performance estimations. The problem with this kind of analysis is the complex process to generating these equations. In this letter, we have developed general equations for decision and sequential processes of CMTC Models, aiming to help researchers to develop steady state availability equations. We also have developed the general equation here termed as Closed Decision Process.

*Introduction:* Continuous Time Markov Chain (CTMC) [1] is one important stochastic tool to describe and analyze systems in many areas of knowledge (Engineering, Biology, Computing, Accountability, etc.). One of the most important analyses of a CTMC model is the steady state availability (SSA). Using (SSA), it is possible to generate equations that can represent the portion of time in which the system stays in each state that can, for example, be used to estimate system performance [2], or even estimate a channel capacity in wireless networks [3]. The problem with this generation is the process necessary to do this (construct the infinitesimal matrix, multiply this matrix with the steady state vector to obtain the system of linear equations, and then solve the system) that depending of the number of states, can require a long time.

In this letter, we firstly describe the general equation of SSA of a state and just then we describe the derivation of general equations to some specific classes of models (sequential and decision processes).

*Theoretical Analysis:*

Theorem 1: Let $S = \{s_1, s_2, \dots, s_n, s_{n+1}, \dots, s_{n+m}\}$ the set of states in a CTMC, $S_n^{'} = \{s_\alpha, s_{\alpha+1}, \dots, s_{\alpha+\beta}\}$ the set of states which the service process led to state $s_n$, $\Gamma = \{\lambda_1, \lambda_2, \dots, \lambda_\beta\}$ the set of arrival rates of $s_n$ and $\Delta = \{\mu_1, \mu_2, \dots, \mu_l\}$ the set of service rates, the general equation of SSA referent to state $s_n$ is :

$$A(s_n) = \left(\frac{\sum_{i=0}^{\beta} \mathrm{A}(S_{n,i}^{'})\Gamma_i}{\sum_{i=0}^{l} \Delta_i}\right) \quad (1)$$

Proof: Given that $P\{T > s + t | T > s\} = P\{T > t\}$ (here referred as memoryless property) [4], and based on Figure 1, it is possible to construct the infinitesimal matrix $Q$ of state transitions as follow:

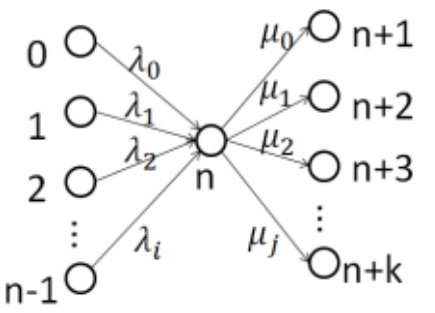

**Fig. 1** *General description of a state in a Markov chain.*

$$Q = \begin{bmatrix} & & & \lambda_0 & & & & \\ & & & \lambda_1 & & & & \\ & & & \vdots & & & & \\ 0 & 0 & \cdots & \theta & \mu_0 & \mu_1 & \cdots & \mu_j \\ & & & 0 & & & & \\ & & & 0 & & & & \\ & & & \vdots & & & & \\ & & & 0 & & & & \end{bmatrix}$$

With $\theta = -(\mu_0 + \mu_1 + \cdots + \mu_j)$. Considering that the columns represent the arrival rates, the rows the service rates, $\Pi$ the set $[0..(n+k)]$ steady probabilities, calculating $Q\Pi = 0$ we have:

$$\lambda_0\pi_0 + \lambda_1\pi_1 + \cdots + \lambda_i\pi_{n-1} - (\mu_0 + \mu_1 + \cdots + \mu_j)\pi_n = 0$$

$$(\mu_0 + \mu_1 + \cdots + \mu_j)\pi_n = \lambda_0\pi_0 + \lambda_1\pi_1 + \cdots + \lambda_i\pi_{n-1}$$

$$\pi_n = \frac{\lambda_0\pi_0 + \lambda_1\pi_1 + \cdots + \lambda_i\pi_{n-1}}{(\mu_0 + \mu_1 + \cdots + \mu_j)} \quad (2)$$

The equation (2) represents the steady state availability general equation related to state n. Based on Equation (2), and using the memoryless property of CTMC, it is possible to derivate the general equations of some particular cases of models. Figure 2(a) represents a case of a decision process; the probabilities of states can be reduced to the state 0 as follow:

$$\pi_1 = \frac{\lambda_0\pi_0}{\lambda_1 + \lambda_2 + \cdots + \lambda_i}; \pi_2 = \frac{\lambda_1\pi_1}{\lambda_{i+1}}; \pi_3 = \frac{\lambda_2\pi_1}{\lambda_{i+2}}; \pi_n = \frac{\lambda_i\pi_1}{\lambda_{i+j}};$$

Considering that $\pi_0 + \pi_1 + \pi_2 + \pi_3 + \cdots + \pi_n + \pi_{n+1} + \cdots + \pi_{n+m} = 1$, we can represent these equations as follow:

$$\pi_0 + \pi_1 + \frac{\lambda_1\pi_1}{\lambda_{i+1}} + \frac{\lambda_2\pi_1}{\lambda_{i+2}} + \cdots + \frac{\lambda_i\pi_1}{\lambda_{i+j}} + \pi_{n+1} + \cdots + \pi_{n+m} = 1;$$

$$\pi_0 + \pi_1\left(1 + \frac{\lambda_1}{\lambda_{i+1}} + \frac{\lambda_2}{\lambda_{i+2}} + \cdots + \frac{\lambda_i}{\lambda_{i+j}}\right) + \pi_{n+1} + \cdots + \pi_{n+m} = 1;$$

$$\pi_0 + \left(\frac{\lambda_0\pi_0}{\lambda_1 + \lambda_2 + \cdots + \lambda_i}\right)\left(1 + \frac{\lambda_1}{\lambda_{i+1}} + \frac{\lambda_2}{\lambda_{i+2}} + \cdots + \frac{\lambda_i}{\lambda_{i+j}}\right) + \pi_{n+1} + \cdots + \pi_{n+m} = 1;$$

$$\pi_0\left(1 + \left(\left(\frac{\lambda_0}{\lambda_1 + \lambda_2 + \cdots + \lambda_i}\right)\left(1 + \frac{\lambda_1}{\lambda_{i+1}} + \frac{\lambda_2}{\lambda_{i+2}} + \cdots + \frac{\lambda_i}{\lambda_{i+j}}\right)\right)\right) + \pi_{n+1} + \cdots + \pi_{n+m} = 1 \quad (3)$$

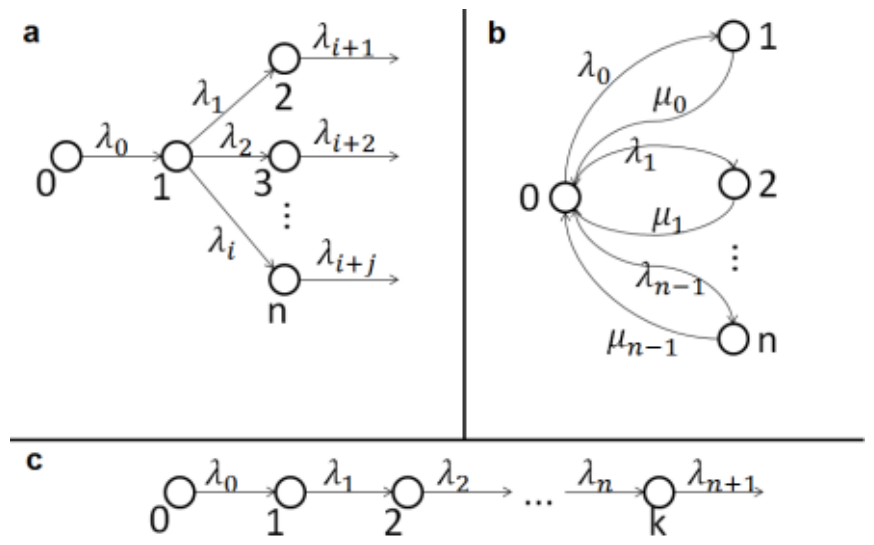

**Fig. 2** *Decision process (a), Closed decision model (b), Sequential process(c).*

The equation (3) shows the derivation of the decision process depicted on Figure 2(a). In this case, the probabilities of states $\pi_1$, $\pi_2$, $\pi_3$, …,$\pi_n$, were reduced in terms of $\pi_0$.

If we consider that transitions $\lambda_{i+1}$, $\lambda_{i+2}$,..., $\lambda_{i+j}$ connect with state 0 closing thus this model we can derive from (3) its general equation as follow:

$$\pi_0\left(1 + \left(\left(\frac{\lambda_0}{\lambda_1 + \lambda_2 + \cdots + \lambda_i}\right)\left(1 + \frac{\lambda_1}{\lambda_{i+1}} + \frac{\lambda_2}{\lambda_{i+2}} + \cdots + \frac{\lambda_i}{\lambda_{i+j}}\right)\right)\right) = 1$$

$$\pi_0 = \frac{1}{\left(1 + \left(\left(\frac{\lambda_0}{\lambda_1 + \lambda_2 + \cdots + \lambda_i}\right)\left(1 + \frac{\lambda_1}{\lambda_{i+1}} + \frac{\lambda_2}{\lambda_{i+2}} + \cdots + \frac{\lambda_i}{\lambda_{i+j}}\right)\right)\right)}$$

$$\pi_0 = \left(1 + \left(\lambda_0\left(\sum_{k=0}^{i} \lambda_k\right)^{-1}\right)\left(1 + \sum_{k=1,y=1}^{i,j} \frac{\lambda_k}{\lambda_{i+y}}\right)\right)^{-1} \quad (3.1)$$

Figure 2(b) presents a type of model that represents one decision system, hereafter termed Closed Decision Model, (an example of this type of model [5]). After we derivate the general equation:

Having:

$$\pi_1 = \frac{\lambda_0\pi_0}{\mu_0};\ \pi_2 = \frac{\lambda_1\pi_0}{\mu_1};\ \pi_n = \frac{\lambda_{n-1}\pi_0}{\mu_{n-1}};$$

And

$$\pi_0 + \pi_1 + \pi_2 + \cdots + \pi_n = 1;$$

We obtain:

$$\pi_0 + \frac{\lambda_0\pi_0}{\mu_0} + \frac{\lambda_1\pi_0}{\mu_1} + \cdots + \frac{\lambda_{n-1}\pi_0}{\mu_{n-1}} = 1;$$

[1] Federal Institute of Education, Science and Technology of Pernambuco.
e-mail: eduardo.vasconcelos@garanhuns.ifpe.edu.br

$$\pi_0\left(1+\frac{\lambda_0}{\mu_0}+\frac{\lambda_1}{\mu_1}+\cdots+\frac{\lambda_{n-1}}{\mu_{n-1}}\right)=1;$$
$$\pi_0=\frac{1}{\left(1+\frac{\lambda_0}{\mu_0}+\frac{\lambda_1}{\mu_1}+\cdots+\frac{\lambda_{n-1}}{\mu_{n-1}}\right)}$$

$$\pi_0=\left(1+\sum_{i=0}^{n-1}\left(\frac{\lambda_i}{\mu_i}\right)\right)^{-1} \quad (4)$$

The equation (4) represents the general equation of model presented on Figure 2b, with regard to state 0.

The Figure 2(c) represents a sequential process; the generic representation of this process can be derived as follow:

$$\pi_1=\frac{\lambda_0\pi_0}{\lambda_1};$$
$$\pi_2=\frac{\lambda_1\pi_1}{\lambda_2}=\frac{\lambda_1\lambda_0\pi_0}{\lambda_2\lambda_1}\Rightarrow\pi_2=\frac{\lambda_0\pi_0}{\lambda_2};$$

So

$$\pi_k=\frac{\lambda_n\pi_{k-1}}{\lambda_{n+1}}=\frac{\lambda_n\lambda_0\pi_0}{\lambda_{n+1}\lambda_n}\Rightarrow\pi_k=\frac{\lambda_0\pi_0}{\lambda_{n+1}};$$

Then, the general equation can be obtained as follow:

$$\pi_0+\pi_1+\pi_2+\cdots+\pi_k+\pi_{k+1}+\cdots+\pi_{k+j}=1;$$
$$\pi_0+\frac{\lambda_0\pi_0}{\lambda_1}+\frac{\lambda_0\pi_0}{\lambda_2}+\cdots+\frac{\lambda_0\pi_0}{\lambda_{n+1}}+\pi_{k+1}+\cdots+\pi_{k+j}=1;$$
$$\pi_0\left(1+\frac{\lambda_0}{\lambda_1}+\frac{\lambda_0}{\lambda_2}+\cdots+\frac{\lambda_0}{\lambda_{n+1}}\right)+\pi_{k+1}+\cdots+\pi_{k+j}=1;$$
$$\pi_0\left(1+\lambda_0\left(\frac{1}{\lambda_1}+\frac{1}{\lambda_2}+\cdots+\frac{1}{\lambda_{n+1}}\right)\right)+\pi_{k+1}+\cdots+\pi_{k+j}=1 \quad (5)$$

As the equation (3), the equation (5) shows that in a sequential process, all probability states ($\pi_1$, $\pi_2$, $\pi_3$, …,$\pi_k$), can be reduced to the sum of the inverse of service rates multiplied by the first arrival rate (state 1 arrival rate) summed to 1.

If we consider that, transition $\lambda_{n+1}$ connects with state 0 on figure 2(c) closing thus this model, we can derive its general equation as follow:

$$\pi_0\left(1+\lambda_0\left(\frac{1}{\lambda_1}+\frac{1}{\lambda_2}+\cdots+\frac{1}{\lambda_{n+1}}\right)\right)=1$$
$$\pi_0=\frac{1}{1+\lambda_0\left(\frac{1}{\lambda_1}+\frac{1}{\lambda_2}+\cdots+\frac{1}{\lambda_{n+1}}\right)}$$
$$\pi_0=\left(1+\lambda_0\left(\sum_{i=0}^{n+1}\frac{1}{\lambda_i}\right)\right)^{-1} \quad (5.1)$$

As $\lambda$ represents the state transition rate and $\lambda=1/E[\tau]$ with $\tau$ representing the sojourn time of the state, if we substitute $\lambda_i$ for $1/E[\tau_i]$ we obtain:

$$\pi_0=\left(1+\left(\frac{1}{E[\tau_0]}\right)\left(\sum_{i=0}^{n+1}\frac{1}{\left(\frac{1}{E[\tau_i]}\right)}\right)\right)^{-1}$$

By solving the above equation algebraically o obtain:

$$\pi_0=E[\tau_0]\left(\sum_{i=0}^{n+1}E[\tau_i]\right)^{-1} \quad (5.2)$$

The equation (5.2) represents the generalization of the ON / OFF process that consists of the ratio between the sojourn time of state i and the sum of all other sojourn times.

*Practical Example:* The Figure 3 represents one closed CTMC that represents a system with one decision and two sequential processes.

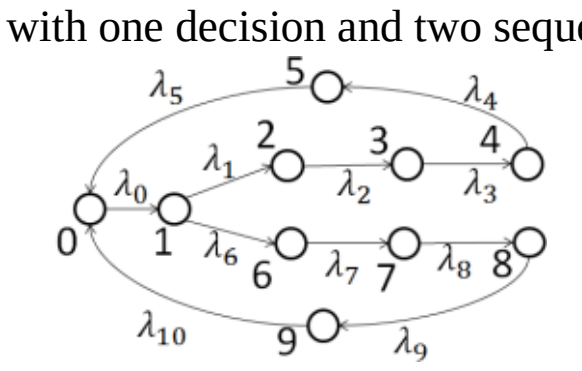

**Fig.** *3 Closed CTMC example.*

If a model is closed, there are only decisions and sequential processes (e.g. as in Figure 3), it is possible to derivate all availability equations based on equations (2), (3) and (4). To do this, it is necessary to derivate the first state equation (In Figure 3, the equation of state 0).
The first step is adding the state 0 on equation as follow:

$$\pi_0+\pi_1+\pi_2+\pi_3+\pi_4+\pi_5+\pi_6+\pi_7+\pi_8+\pi_9=1 \quad (6)$$

Secondly, we add the state 1, on the equation (6). As all states can be expressed in terms of $\pi_0$, we can directly add the arrival and service rates on equation. As state 1 is one decision process, we have:

$$\pi_0\left(1+\frac{\lambda_0}{\lambda_1+\lambda_2}\right)+\pi_2+\pi_3+\pi_4+\pi_5+\pi_6+\pi_7+\pi_8+\pi_9=1 \quad (7)$$

Considering the memoryless property, we can assert that the others states will depend on state 1, so to the others states we have:

$$\pi_0\left(1+\frac{\lambda_0}{\lambda_1+\lambda_2}\left(1+\frac{\lambda_1}{\lambda_2}\right)\right)+\pi_3+\pi_4+\pi_5+\pi_6+\pi_7+\pi_8+\pi_9=1 \quad (8)$$

As from state 1, we have two sequential processes; the rest of the equation is derived as follow:

$$\pi_0\left(1+\frac{\lambda_0}{\lambda_1+\lambda_2}\left(1+\frac{\lambda_1}{\lambda_2}+\frac{\lambda_1}{\lambda_3}+\frac{\lambda_1}{\lambda_4}+\frac{\lambda_1}{\lambda_5}+\frac{\lambda_6}{\lambda_7}+\frac{\lambda_6}{\lambda_8}+\frac{\lambda_6}{\lambda_9}+\frac{\lambda_6}{\lambda_{10}}\right)\right)=1$$
$$\pi_0=\frac{1}{1+\frac{\lambda_0}{\lambda_1+\lambda_2}\left(1+\frac{\lambda_1}{\lambda_2}+\frac{\lambda_1}{\lambda_3}+\frac{\lambda_1}{\lambda_4}+\frac{\lambda_1}{\lambda_5}+\frac{\lambda_6}{\lambda_7}+\frac{\lambda_6}{\lambda_8}+\frac{\lambda_6}{\lambda_9}+\frac{\lambda_6}{\lambda_{10}}\right)} \quad (9)$$

So, if we want to derivate the equation of state 4, we need only to describe the equation in terms of $\pi_0$:

$$\pi_4=\frac{\lambda_1\pi_1}{\lambda_4};\ \pi_1=\frac{\lambda_0\pi_0}{\lambda_1+\lambda_6};$$
$$\pi_4=\frac{\lambda_1\lambda_0\pi_0}{\lambda_4(\lambda_1+\lambda_6)} \quad (10)$$

*Conclusion:* In this letter, we have presented a theoretical analysis of Continuous Time Markov Chain. With this analysis we have derived some general equation of steady state availability of some types of models. These equations and analysis can help researchers to derivate availability equations more easily, because there is no need to create the infinitesimal matrix, sequential and decision processes can be derived more quickly, decreasing thus the amount of linear equations that must be derived.

**References**

1 Marsan, M. J. ‘Modelling with Generalized Stochastic Petri Nets’ Wiley Series in Parallel Computing John Wiley and Sons, 1995.

2 Charki, A.; Bigaud, D., "Availability Estimation of a Photovoltaic System," Reliability and Maintainability Symposium (RAMS), 2013 Proceedings - Annual , vol., no., pp.1,5, 28-31 Jan. 2013.

3 Vasconcelos, E.; Dias, K.L.; Cunha, P.R.F.; Cavalcanti, D.; Cordeiro, C., "Joint modelling of medium access and primary/secondary users for cognitive radios through Markov chain," Electronics Letters , vol.49, no.15, pp.,, July 18 2013.

4 Norris, J. R. Markov chains. Reprint of 1997 original. Cambridge Series in Statistical and Probabilistic Mathematics, 2. Cambridge University Press, Cambridge, 1998. xvi+237 pp. ISBN: 0-521-48181-3.

5 Xiaorong Zhu; Lianfeng Shen; Yum, T.-S.P., "Analysis of Cognitive Radio Spectrum Access with Optimal Channel Reservation," Communications Letters, IEEE , vol.11, no.4, pp.304,306, April 2007.